\pdfoutput=1

\documentclass[3p,times,twocolumn]{elsarticle}

\usepackage{ecrc}

\volume{00}

\firstpage{1}

\journalname{Nuclear Physics B Proceedings Supplement}

\runauth{}

\jid{nppp}

\jnltitlelogo{Nuclear Physics B Proceedings Supplement}

\usepackage{amssymb}

\usepackage[figuresright]{rotating}

\begin{document}

\begin{frontmatter}



\dochead{}

\title{Chemically non-equilibrated QGP and thermal photon elliptic flow}


\author{Akihiko Monnai}

\address{RIKEN BNL Research Center, Brookhaven National Laboratory, Upton, NY 11973, USA}

\begin{abstract}
It has been discovered in recent heavy-ion experiments that elliptic and triangular flow of direct photons are underpredicted by most hydrodynamic models. I discuss possible enhancement mechanisms based on late chemical equilibration of the QGP and in-medium modification of parton distributions. Numerical hydrodynamic analyses indicate that they suppress early photon emission and visibly enhance thermal photon elliptic flow.  
\end{abstract}


\end{frontmatter}


\section{Introduction}
\label{sec:1}

Flow harmonics of particle spectra ($v_n$) are good observables to quantify the magnitude of interaction in a hot medium created in ultrarelativistic nuclear colliders \cite{Ollitrault:1992bk,Poskanzer:1998yz}. They can be defined as
\begin{equation}
\label{eq:vngamma}
v_n (p_T,y) = \frac{\int_0 ^{2\pi} d\phi_p \cos (n\phi_p - \Psi_n) \frac{dN}{d\phi_p p_Tdp_T dy}}{\int_0 ^{2\pi} d\phi_p  \frac{dN}{d\phi_p p_Tdp_T dy}} ,
\end{equation}
where $N$ is the particle number, $p_T$ is the transverse momentum, $y$ is the rapidity, $\phi_p$ is the azimuthal momentum angle and $\Psi_n$ is the reference angle for the $n$-th harmonics.
Experimental data at BNL Relativistic Heavy Ion Collider (RHIC) and CERN Large Hadron Collider (LHC) indicate that geometrical anisotropy of the overlapping region of colliding nuclei is clearly mapped onto hadronic momentum anisotropy, leading to the notion of strongly-coupled quark-gluon plasma (sQGP). Relativistic viscous hydrodynamic models are known to provide a quantitative description of the particle spectra and differential flow harmonics \cite{Schenke:2010rr}, and are also studied extensively for the analyses of the beam energy scan and the small system programs at RHIC. 

One can similarly define the flow harmonics of direct photons. Here the direct photons are defined as the sum of prompt photons and thermal photons, where the former are created in the initial hard processes and the latter are emitted from the medium. Although the thermal photons do not interact strongly with the medium, they can inherit some momentum anisotropy from the medium. Recent direct photons measurements have revealed that its elliptic flow $v_2$ and triangular flow $v_3$ are as large as hadronic counterparts \cite{Adare:2011zr,Lohner:2012ct,Mizuno:2014via}. So far hydrodynamic models fail to describe this as they tend to predict much smaller values owing to the contributions of early photons with little azimuthal momentum anisotropy. This has been recognized as ``photon puzzle", and many theoretical attempts have been made to understand this intriguing problem. 

In this study, I consider two possible mechanisms which lead to enhancement of the photon flow harmonics. First, the glasma, which follows from the color glass condensate (CGC) \cite{McLerran:1993ni,McLerran:1993ka}, has many high-momentum gluons before the QGP is formed \cite{Monnai:2014xya}. Chemical equilibration can be slower than the thermalization because the former is driven by inelastic scatterings while the latter by both elastic and inelastic scatterings. This implies that quarks are not fully produced at the onset of hydrodynamic stage \cite{Monnai:2014kqa}. Second, the QGP is not a free gas as suggested by the thermodynamic results of lattice QCD \cite{Karsch:2000ps}, and the equilibrium distributions should be modified from Fermi-Dirac or Bose-Einstein ones \cite{Monnai:2015qha}. The former mechanism leads to the suppression of photons in early hydrodynamic stages and the latter one to the reduction of photons from high-temperature regions. They both enhance the photon flow harmonics because relative contributions of the thermal photons emitted from the fluid elements with fully-developed momentum anisotropy become effectively large. I also show numerically that thermal photon $v_2$ is visibly enhanced by those mechanisms.

\section{Quark chemical equilibration}
\label{sec:2}

The heavy-ion system at high energies is described as saturated gluons at earliest times. Chemical equilibration of quark components can take longer than thermalization as elastic scattering processes do not contribute to the former, even though they are often assumed to occur simultaneously in hydrodynamic analyses. This is also a good analogy to the difference in the concepts of thermal and chemical freeze-outs. One can expect suppression of thermal photon emission at the beginning of the hydrodynamic stage because quarks are required for photon production. Since flow anisotropy develops during the time evolution, the mechanism would effectively enhance the photon flow harmonics. 

Here I model chemical equilibration by introducing rate equations, which should be solved with energy-momentum conservation $\partial_\mu T^{\mu \nu} = 0$:
\begin{eqnarray}
\partial_\mu N_q^\mu &=& 2 r_b n_g - 2 r_b \frac{n_g^\mathrm{eq}}{(n_q^\mathrm{eq})^2} n_q^2 , \label{eq:quark} \\
\partial_\mu N_g^\mu &=& (r_a - r_b) n_g - r_a \frac{1}{n_g^\mathrm{eq}} n_g^2 + r_b \frac{n_g^\mathrm{eq}}{(n_q^\mathrm{eq})^2} n_q^2 , \nonumber \\
&+& r_c n_q - r_c \frac{1}{n_g^\mathrm{eq}} n_q n_g \label{eq:gluon} .
\end{eqnarray}
Here $N_q^\mu = n_q u^\mu$ and $N_g^\mu = n_g u^\mu$ are the quark and the gluon number currents. The quark-antiquark degrees of freedom is included in the quark number by assuming the vanishing limit of baryon chemical potential. Viscosity and diffusion are not considered here. $n_q^\mathrm{eq}$ and $n_g^\mathrm{eq}$ are the quark and the gluon densities in equilibrium, which are determined using a quasi-particle description of lattice QCD results in Sec.~\ref{sec:3}. Gluon splitting, quark pair production and gluon emission of quarks are characterized by the reaction rates $r_a$, $r_b$ and $r_c$, respectively. The recombination processes are introduced so the rate equations become relaxation-like equations. The off-shell patrons are assumed to be prepared by in-medium interaction, and thermalization to be achieved immediately after the number changing processes.

It should be noted that the quark number is changed only by the quark pair production and annihilation here. This implies that the quark chemical equilibration is characterized by the time scale of $\tau_\mathrm{chem} \sim 1/r_b$. The initial conditions for the rate equations are chosen as $n_q = 0$ and $n_g = n_g^\mathrm{eq} + n_q^\mathrm{eq}/2$ to simulate initial overpopulation of gluons. The reaction rates are parametrized as
$r_a = c_a T$, $r_b = c_b T$ and $r_c = c_c T$ with free dimensionless parameters $c$.

The concept of partons are valid at most above the crossover temperature. The system is assumed to be in complete chemical equilibrium in the hadronic phase.

\section{In-medium parton distributions}
\label{sec:3}

The equation of state (EoS) of lattice QCD implies that the QGP at heavy-ion temperatures is not described as a parton gas. On the other hand, Fermi-Dirac or Bose-Einstein distributions are inconsistently employed in most heavy-ion models. Since the QGP photon emission rate is a functional of the parton distribution functions, it is important to take the effects of in-medium corrections into account. The effective number of the degrees of freedom is smaller than that of the free gas in the QGP phase. Thus, similar to the chemical equilibration scenario, flow harmonics of photons will have an additional boost owing to the suppression of early photon emission. Note that the in-medium corrections in the hadronic phase is small because the EoS is well described by hadronic resonance gas. 

The quasi-particle model is formulated as follows. I introduce the effective energy density $\omega^i = (p^2 + m_i^2)^{1/2} + W^i_\mathrm{eff}$, where $W^i_\mathrm{eff}$ represents the in-medium correction, to construct an effective distribution function:
\begin{equation}
f_\mathrm{eff}^i = \frac{1}{\exp{(\omega_i /T)}\pm1} .
\end{equation}
The partition function reads
\begin{equation}
\frac{\ln Z_i}{V} = \pm \int \frac{g_i d^3p}{(2\pi)^3} \ln{\bigg[1\pm \exp{\bigg(- \frac{\omega_i}{T}\bigg) } \bigg]} - \frac{\Phi_i}{T}, 
\end{equation}
where $V$ is the volume, $g_i$ is the degeneracy and $\Phi_i$ is the background field contribution dependent on the temperature. The thermodynamic consistency condition \cite{Biro:2001ug} 
\begin{eqnarray}
\frac{\partial \Phi_i}{\partial T}\Big|_{\mu_B} = - \int \frac{g_i d^3p}{(2\pi)^3} \frac{\partial \omega_i}{\partial T}\Big|_{\mu_B} f_\mathrm{eff}^i , \label{eq:tcc}
\end{eqnarray}
relates $\Phi_i$ and $\omega_i$, where the baryon chemical potential $\mu_B = 0$.
The standard thermodynamic relations lead to 
\begin{eqnarray}
e &=& - \frac{1}{V} \sum_i \frac{\partial \ln Z_i}{\partial \beta}\Big|_{\alpha_B} \nonumber \\
&=& \sum_i \int \frac{g_i d^3p}{(2\pi)^3} \omega_i f^i_\mathrm{eff} + \Phi, \label{eq:e} \\
P &=& \frac{1}{V} \sum_i T \ln Z_i \nonumber \\
&=& \pm T \sum_i \int \frac{g_i d^3p}{(2\pi)^3} \ln{\bigg[1\pm \exp \bigg(-\frac{\omega_i}{T} \bigg) \bigg]} - \Phi, \label{eq:P} \nonumber \\
\end{eqnarray}
where $\Phi = \sum_i \Phi_i$ and $\alpha_B = \mu_B/T = 0$.

The effective correction to the energy $W^i_\mathrm{eff}$ is determined so that $e$ and $P$ of lattice QCD calculations \cite{Borsanyi:2013bia} are reproduced. Here it is simply assumed to be dependent on the temperature and not on particle species because of the lack of additional constraints. 

\begin{figure}[tb]
\includegraphics[width=2.6in]{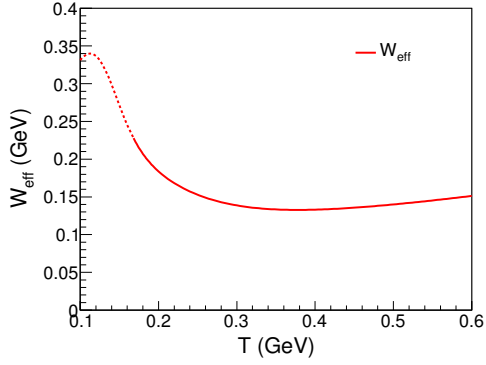}
\caption{The effective correction to the energy $W_\mathrm{eff}$. The dotted line denotes the temperature region below 0.17 GeV.}
\label{fig:1}
\end{figure}

\begin{figure}[tb]
\includegraphics[width=2.6in]{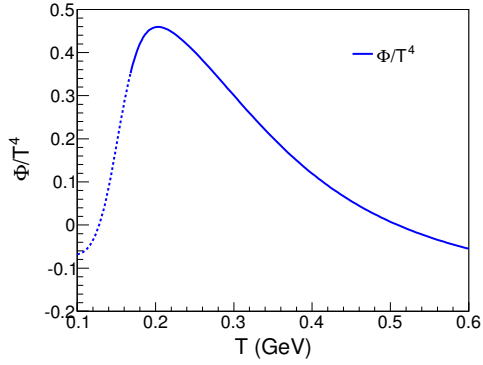}
\caption{The background field contribution $\Phi/T^4$. The dotted line denotes the temperature region below 0.17 GeV. }
\label{fig:2}
\end{figure}

Figure~\ref{fig:1} shows $W_\mathrm{eff}$ as a function of the temperature. One can see that the correction is comparable to the temperature and thus would be non-negligible. The background contribution $\Phi/T^4$ is shown in Fig.~\ref{fig:2} as a function of the temperature using Eq.~(\ref{eq:tcc}). It is note-worthy that $\Phi$ is relatively small compared to the energy density and the pressure.

\section{Hydrodynamic analyses}
\label{sec:4}

The thermal photon emission rate is calculated based on the one for the QGP \cite{Strickland:1994rf, Srivastava:1996qd} and the hadron \cite{Turbide:2003si, Arleo:2004gn} phases. The former reads, as functions of the effective fugacities $\lambda_q$ and $\lambda_g$,
\begin{eqnarray}
E\frac{dR}{d^3p} &=& \frac{5 \alpha \alpha_s}{9\pi^2} T^2 \exp(-E/T) \nonumber \\
&\times& \bigg\{ \lambda_q \lambda_g \bigg[ \log \bigg( \frac{4ET}{k_c^2} \bigg) + \frac{1}{2} - \gamma \bigg] \nonumber \\
&+& \lambda_q^2 \bigg[ \log \bigg( \frac{4ET}{k_c^2} \bigg) - 1 - \gamma \bigg] \bigg\}. \label{eq:rate} 
\end{eqnarray}
Here $\lambda_q = (n_q/n_q^\mathrm{eq}) \exp{(-W_\mathrm{eff}/T)}$ and $\lambda_g = (n_g/n_g^\mathrm{eq}) \exp{(-W_\mathrm{eff}/T)}$. $\gamma$ is Euler's constant and $k_c^2 = g^2T^2/6$ is the infra-red cut-off momentum. The QGP photon emission rate is smoothly interpolated with the hadronic one using a hyperbolic function at the crossover temperature $T_c = 0.17$ GeV \cite{Monnai:2015qha}.

The space-time evolution of the medium is estimated using a (2+1)-dimensional ideal hydrodynamic model \cite{Monnai:2014kqa}. A Monte-Carlo Glauber model \cite{Miller:2007ri} is employed for obtaining the event-averaged initial condition for 200 GeV Au-Au collisions at the impact parameter $b = 6$ fm. 
The equation of state is the same as the one used in the quasi-particle model discussed in Sec.~\ref{sec:3}. The initial time is $\tau_0 = 0.4$ fm/$c$. The thermal photon emission above $T_f = 0.13$ GeV is taken into account.

\begin{figure}[tb]
\includegraphics[width=2.6in]{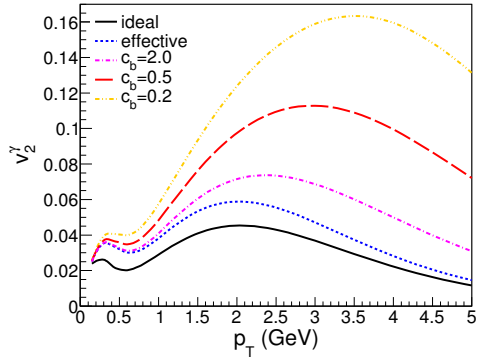}
\caption{Thermal photon $v_2^\gamma (p_T)$. The solid and dotted lines correspond to the results with ideal and effective distribution in chemical equilibrium. The dash-dotted, dashed and dash-double-dotted lines are for the results with different chemical equilibration rates.}
\label{fig:3}
\end{figure}

Figure~\ref{fig:3} shows differential thermal photon elliptic flow $v_2^\gamma$ with and without in-medium effective corrections, and those with and without chemical equilibration processes in addition to the effective corrections. $v_2^\gamma$ is visibly enhanced by both mechanisms. The difference between ideal and effective distributions can be seen at lower $p_T$. The reaction rate parameters $c_a = c_c = 1.5$ and $c_b = 0.2, 0.5$ and $2.0$ are chosen for the chemically non-equilibrated results. Slower chemical equilibration leads to larger $v_2$ because early photon emission is suppressed more. 

\begin{figure}[tb]
\includegraphics[width=2.6in]{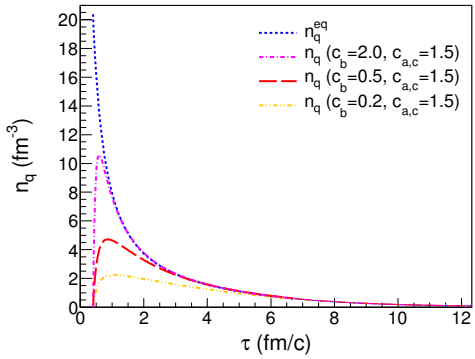}
\caption{Time evolution of $n_q$ at $x=y=0$~fm. See the caption of Fig.~\ref{fig:3} for the line types. The thin lines denote the region below 0.17~GeV.}
\label{fig:4}
\end{figure}

The time evolution of the quark number density at the center of the collision is shown in Fig.~\ref{fig:4}. Numerical estimations agree with the observation that $c_b = 0.2, 0.5$ and $2.0$ roughly correspond to the chemical relaxation times $\tau_\mathrm{chem} \sim 5, 2$ and $0.5$ fm/$c$, respectively, when the average medium temperature is 0.2 GeV. 

\begin{figure}[tb]
\includegraphics[width=2.6in]{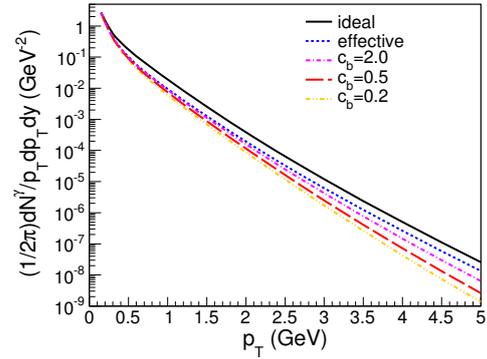}
\caption{Thermal photon $p_T$ spectra. See the caption of Fig.~\ref{fig:3} for the line types.}
\label{fig:5}
\end{figure}

Figure~\ref{fig:5} is the plot for thermal photon $p_T$ spectra. The suppression of photons leads to the reduction of the total photon yield in both cases. It is note-worthy that the effects of chemical equilibration appear mainly at high $p_T$ while those of in-medium effective corrections are visible in a wide $p_T$ range. This implies that the photons with lower $p_T$ are produced at later times.

\section{Summary and outlook}
\label{sec:6}

The effects of quark chemical equilibration and effective corrections to the parton distributions on the elliptic flow and particle spectra of thermal photons are investigated. The former leads to the suppression of early photons and the latter to the suppression of photons from hot fluid elements. While the effects of chemical equilibration may be larger, they both enhance thermal photon $v_2^\gamma$, which is in qualitative agreement with the recent experimental observations. On the other hand, $p_T$ spectrum is suppressed. It can be seen that those non-exotic mechanisms play important roles in understanding the direct photons in heavy-ion collisions.

Future prospects include estimation of prompt photon contributions and calculations of higher order flow harmonics. Introduction of chemical non-equilibrated EoS \cite{Gelis:2004ep} is also important for more quantitative analyses.

\section*{Acknowledgment}
I would like to thank Y.~Akiba, M.~Asakawa, M.~Kitazawa, L. McLerran, K.~Morita, B.~M\"{u}ller, A.~Ohnishi and B.~Schenke for valuable comments. The work of A.M. is supported by the RIKEN Special Postdoctoral Researcher program. Some of the results are calculated using RIKEN Integrated Cluster of Clusters (RICC).




\nocite{*}
\bibliographystyle{elsarticle-num}
\bibliography{martin}



\end{document}